# Combination of Lidar Elevations, Bathymetric Data, and Urban Infrastructure in a Sub-Grid Model for Predicting Inundation in New York City during Hurricane Sandy

Jon Derek Loftis [a,*], Harry V. Wang [a], Stuart E. Hamilton [b], and David R. Forrest [a]

[a] Department of Physical Sciences, Virginia Institute of Marine Science, College of William and Mary, P.O. Box 1375, Gloucester Point, VA 23062, USA
[b] Department of Geography and Geosciences, Salisbury University, Salisbury, MD 21801, USA

[*] Corresponding Author. Tel.: 8046847876.  Email address: jdloftis@vims.edu

| ARTICLE INFO | ABSTRACT |
|---|---|
| *Highlights:*<br>• Features a sub-grid model specifically designed to model flooding in urban environments.<br>• Hurricane Sandy in New York City is used as an inundation case study.<br>• New method presented for integrating lidar topography, bathymetry, and buildings into a DEM and sub-grid model.<br>• Incorporates highway underpass elevations to minimize vertical occlusion and artificial impediments to storm surge for improved results.<br>• Verified observation data from USGS and FEMA was utilized to characterize spatial results.<br><br>*Keywords:*<br>GIS; Urban Flooding; Hydrodynamic Modeling; Storm Surge; Buildings; Underpass | In this paper, we present the geospatial methods in conjunction with results of a newly developed storm surge and sub-grid inundation model which was applied in New York City during Hurricane Sandy in 2012. Sub-grid modeling takes a novel approach for partial wetting and drying within grid cells, eschewing the conventional hydrodynamic modeling method by nesting a sub-grid containing high-resolution lidar topography and fine scale bathymetry within each computational grid cell. In doing so, the sub-grid modeling method is heavily dependent on building and street configuration provided by the DEM. These prominent ultra-urban infrastructural features are often avoided or otherwise loosely accounted for in typical non-sub-grid modeling approaches, yet inherently accounted for using the sub-grid approach to efficiently simulate street-level inundation. The results of spatial comparisons between the sub-grid model and FEMA's maximum inundation extents in New York City yielded an unparalleled absolute mean distance difference of 38m and an average of 75% areal spatial match. An in-depth error analysis reveals that the modeled extent contour is well correlated with the FEMA's extent contour in most areas, except in several distinct areas where differences in special features cause significant de-correlations between the two contours. Examples of these errors were found to be primarily attributed to lack of building representation in the New Jersey region of the model grid, occluded highway underpasses artificially blocking fluid flow, and DEM source differences between the model and FEMA. Accurate representation of these urban infrastructural features is critical in terms of sub-grid modeling, because it uniquely affects the fluid flux through each grid cell side, which ultimately determines the water depth and extent of flooding via distribution of water volume within each grid cell. Incorporation of buildings and highway underpasses allow for the model to improve overall absolute mean distance error metrics from 38m to 32m (a 15.8% improvement) and area comparisons from 75% spatial match to 80% with minimal additional effort. This demonstrates the importance of accurately incorporating detailed topographic features in a DEM for better prediction of urban inundation using sub-grid modeling. |

## 1. Introduction

Numerous storm surge models have been developed and applied along the U.S. East Coast, and they vary based upon grid type (structured/unstructured) or upon the numerical schemes used (implicit, semi-implicit, explicit), with examples including SLOSH, ADCIRC, FVCOM, ROMS, SELFE, and others (Jelesnianski, et al., 1992; Westerink et al., 1994; Zhang et al., 2008). This study utilizes a highly-resolved, sub-grid inundation model (Casulli and Stelling, 2011), which makes use of a lidar-derived digital elevation model (DEM) specifically produced for the New York City metropolitan area to address the extent, timing, and depth of inundation at the street-level during 2012 Hurricane Sandy.

Given the variety of densely-compacted multi-scale topographic features prevalent in an ultra-urban setting, a coarse computational grid cannot be efficiently scaled to incorporate all the unique objects, features, and scales therein. Thus, an efficient and plausible approach is to sub-divide the various scales and dimensions of buildings and streets down to the smallest basic unit of the sub-grid cell. Resolving multiple features at the sub-grid scale permits calculation of form drag posed by those features and skin friction as the shallow surge propagates through city streets during a flooding event. Furthermore, by utilizing a non-linear solver and the conveyance formulation for calculating flow resistance, it effectively improves model accuracy to the street-level scale without the high computational cost of simulation on a fully-fledged high-resolution hydrodynamic model grid (Casulli and Zanolli, 2012; Wang et al., 2014).

Sub-grid modeling is a multi-scale approach to hydrodynamic modeling by which water level elevations on a sub-grid level can be obtained through the combination of water levels and velocities efficiently calculated at the coarse computational grid, the discretized bathymetric depths, and local friction parameters, without utilizing the excessive computing resources required to solve the full set of shallow water equations (Casulli and Stelling, 2011). Sub-grid technology essentially allows velocity to be determined efficiently at the sub-grid scale. This salient feature enables coastal flooding to be addressed in a single cross-scale model from the ocean to upstream river channels without overly refining the grid resolution. To this end, high-resolution topobathymetric DEMs were developed using GIS from lidar-derived point clouds for incorporation into a sub-grid model, for research into the plethora of practical applications related to inundation.

The most recent relevant application for combining lidar-derived elevations, bathymetric data, and structures in a multi-scale model to predict urban flooding would be to model Hurricane Sandy in New York City. This study will use 2012 Hurricane Sandy as a key case study to address the challenges associated with modeling storm surge and tide interaction and their associated impacts on spatial inundation extents. Additionally, employing new technology using sub-grid modeling coupled with high-resolution lidar-derived topography, and the inclusion of complex building infrastructure to simulate inundation observed in the urban environment surrounding the New York Harbor region will permit a detailed local flooding analysis for Hurricane Sandy.

Hurricane Sandy is the second-costliest hurricane on record (after 2005 Hurricane Katrina) to make landfall in the United States. While only a Category 1 storm on the Saffir-Simpson scale when it made landfall in Atlantic City, NJ, Hurricane Sandy was directly responsible for 73 deaths, and amounted to more than $65 billion dollars in assessed damages in the United States (NOAA NCDC, 2013; Smith and Katz, 2013). Hurricane Sandy directly impacted 24 states, especially all coastal states across the Mid-Atlantic Bight, with the most severe damage accounted for in New Jersey and New York. Hurricane Sandy made landfall along the New Jersey coast, south of the New York Harbor, on October 29th, with dual storm surges approaching from the south through the New York Bay and from the east propagating via the Long Island Sound (NOAA Service Assessment, 2012; NOAA NCDC, 2013). New York City, NY, along with Jersey City, NJ, and Hoboken, NJ, were heavily impacted by the effect of the storm surge bottlenecking up the Hudson River and East River systems, with the storm surge flooding streets, tunnels, and subway lines; effectively cutting electrical power, as sub-surface electrical infrastructure became flooded via transit tunnels throughout the city (Smith and Katz, 2013).

Water from Hurricane Sandy's storm surge flooded into New York City, encountering complex, developed land surfaces characterized by a wide range of unique features ranging from waterfront berms, streets, railroads, parks, highways, subway stations, and bridges, along with a variety of building types. High-resolution hydrodynamic models are needed to appropriately consider the impact of these local features into the prediction of maximum storm surge extents. Even with the ample computing resources available today, it is still insufficient to model all complex topographic features at the individual building scale or at street-level resolution. Recent research demonstrates that, provided lidar data of topographic heights and sufficient bathymetric water depths, both of which can be collected with very high resolution, incorporation of detailed elevation measurement within a coarse grid model can be used to further improve model accuracy (Casulli, 2009; Loftis et al., 2013). This is the emerging consensus for the multi-scale sub-grid modeling approach.

A multi-faceted approach will be used to address spatial verification of the inundation extent of Hurricane Sandy's storm surge predicted by the street-level inundation model. These methods will include: point-to-point comparisons to validate flood water depths, along with multiple distance comparisons between FEMA's maximum flood extent map and modeled street-level inundation results, and separate area comparisons along the New York and New Jersey banks of the Hudson River, the East River, and the Harlem River. Additionally, a suite of time series observations from NOAA tide gauges and USGS overland rapid-deployment gauges were utilized for calibration and validation of the model results, previously published in Wang et al., 2014 (NOAA Tides and Currents, 2012; McCallum et al., 2013).

Subsequent sections discuss the study area surrounding New York City, and utilizing GIS to join recent lidar-derived topographic measurements with recent bathymetric measurements from several NOAA surveys, and the addition of buildings to the DEM. Section 3 introduces the methodology for development of an effective sub-grid model for predicting spatial inundation extents using lidar-derived topography with building integration in an ultra-urban environment. Section 4 presents the observation data used to validate model predictions along with GIS post-processing methods for computing statistics used in spatial comparisons. Section 5 presents results of a rigorously conducted series of spatial analyses of inundation distance and area comparisons for quality verification using several reputable observations from U.S. government agencies followed by concluding remarks in section 6.

**2. Study area and DEM development**

This modeling effort encapsulates coastal regions of New York City and New Jersey along the New York Bay, Hudson River, East River, and Harlem River, with dimensions of 37 km north to south, and 29 km west to east (Figure 1). The study region is bounded in the west by the NOAA-operated gauge at Bergen Point along the Kill van Kull, east to where the Long Island Sound meets the East River near the NOAA gauge at Kings Point, north to Yonkers, and South to Coney Island. New York City is the preeminent commercial trade capital globally, notably housing the New York Stock Exchange, and a substantial population of greater than 8.3 million residents with a population density of 10,725 people per km² (United States Census Bureau, 2012). Some of the world's highest value real estate can be found in this region alongside a major port and numerous other sites of national importance. The setup and design of the DEM to be used with the New York City sub-grid includes multiple topographic and bathymetric data sources with the addition of buildings for the metropolitan area of New York City. All data utilized were projected in UTM zone 18N (EPSG:26918) for use in spatial analyses.

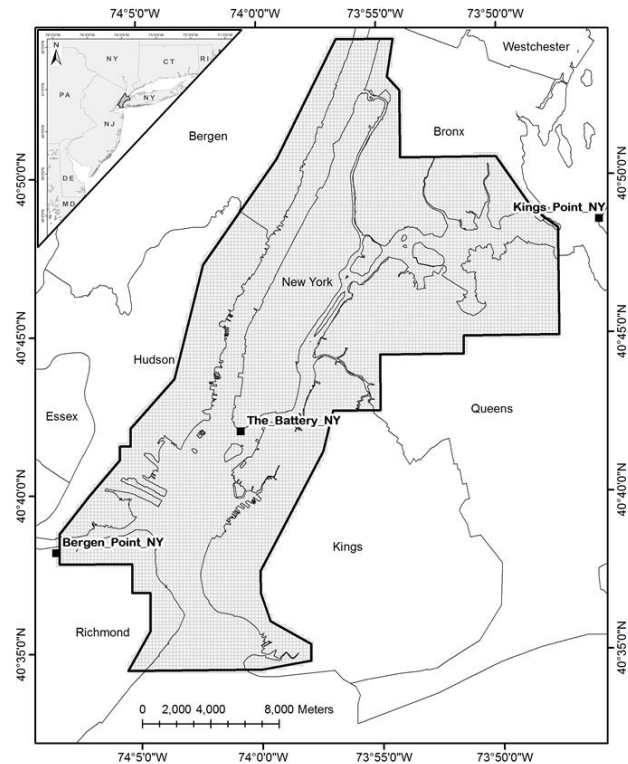

**Figure 1.** Study area including locations of 3 NOAA tide gauges.

*2.1 Topography data*

The USGS National Elevation Dataset (NED) was the primary source of topography utilizing the 1/9 arc-second data where available, supplemented with 1/3 arc-second data where no higher resolution was available (Gesch et al., 2002). These data are derived from lidar point data flown in December 2006 through February 2007 as part of a USGS update initiative for the NED to incorporate higher resolution topographic data (Gesch, 2007). The data for New York City and New Jersey were downloaded and merged into a singular topographic dataset and resampled from either ≈3m or ≈10m to 5m resolution.

Modeling underpasses and overpasses is a major challenge in urban environments (Loftis, 2014). Within the area of New York City covered in this study, a total of 810 street segments are classified as bridges. The creation of a surface model from lidar topography results in a surface that most often reflects the elevation of the higher portion of the bridge, and this, in turn, prohibits water from flowing under the road or rail bridge. We overcame this hindrance by applying the lower height of the overpass at bridge locations. To achieve this, we obtained all roads labelled as bridges from Open Street Map and a commercial vendor. We extracted all roads that intersected the location of the bridges (these are the roads that are under the bridges). We then split these road segments either side of the bridge and extracted the lowest elevation from the surface DEM of the intersecting road segment. This linear path was expanded to account for the width of the road and DEM was coded with the lower elevation in this area. This allows for flooding to pass under bridges into areas that would otherwise have received no model flooding due to the overpass acting as a flood barrier in the DEM.

*2.2 Bathymetry data*

Coastal relief data were downloaded as an ASCII file at 90m resolution from NOAA's National Geophysical Data Center and imported into ArcGIS as the base bathymetry DEM. Higher resolution (10m) NOAA digital bathymetric survey data were collected where available and assimilated into the bathymetric DEM while assuring elevation symmetry along the seams. The following digital data surveys were collected from NOAA: H11600 collected in 2006 along the New York Bay and Verrazano Narrows in the south central area of the sub-grid domain, H11353 collected in 2004 along the East River, and H11395 gathered in 2006 along the Hudson River adjacent to Manhattan Island (NOAA NOS, 2006). The merged bathymetry data were then reprojected to EPSG:26918 and resampled to 5m resolution to match the resolution of the merged topographic DEM. The bathymetric data were then converted to ASCII format for input into the sub-grid generation software.

*2.3 Building integration*

Vector building footprints with embedded building heights were obtained from the NYC DOITT GIS repository (New York City Buildings, 2013). The five boroughs of New York City were merged from five vector datasets into one and were reprojected to use the same geographic projection used for the topographic and bathymetric data. No preprocessed vector building footprints were freely available for the New Jersey portion of the study area, and these areas will be used to illustrate the value of building integration into the sub-grid model. Using the polygon building heights layer, a template geotiff raster of buildings was resampled to 5m resolution using the building heights field as the elevation above the local topography. The geotiff output for the building layer DEM was exported via ASCII format for compatible use with Janet, the sub-grid generation software.

Typically, buildings and other prominent features in ultra-urban environments are neglected due to the high computational cost incurred when resolving numerous individual features in the hydrodynamic model grid. However, this paper emphasizes the usage of a sub-grid model to efficiently resolve fine scale buildings to more effectively model street-level inundation. Each of these DEMs for buildings, bathymetry, and bare-earth lidar-derived topography measurements will be combined in the hydrodynamic model sub-grid in the following section.

## 3. Sub-grid model description and setup

A hydrodynamic sub-grid model is utilized in this study to simulate storm surge and inundation caused by 2012 Hurricane Sandy. The numerical algorithms of the sub-grid model are both robust, and computationally efficient (Casulli and Walters, 2000; Casulli and Zanolli, 2002; Casulli, 2009; Casulli and Stelling, 2011). More detailed model descriptions can be found in the previous references. Recent technological advancements in cross-scale modeling methods allow for the use of a sub-grid mesh embedded within each base computational grid element with an inherent numerical scheme capable of partial wetting and drying. The sub-grid model possesses numerous other valuable properties including: high-order numerical accuracy, global and local mass conservation, and unconditional stability due to its computationally semi-implicit scheme (Casulli and Stelling, 2011). Greatest numerical accuracy is achieved when a uniform grid, comprised of uniform quadrilaterals (like squares) or equilateral triangles, is used. For this reason, many of the grids developed using lidar-derived data have been scaled to square grids congruent to the native resolution of the topographic data contained in the DEM. The sub-grid model grid utilized to model 2012 Hurricane Sandy in the New York Harbor region makes use of a 200m base grid with a 40×40 nested 5m sub-grid within each grid cell (Figure 2). The grid is comprised of 11,959 nodes, 23,559 sides, and 11,601 elements, covering an area of 29×37km, and translating to 4,496,833 sub-grid cells at 5m resolution.

*3.1 Sub-grid model development*

Using the sub-grid generation software, Janet v.2.9.36 (Lippert, 2010), domain extent polygons (shown as black lines in Figure 1) were imported to be the template for the sub-grid boundary. The polygon editor was utilized to copy the imported boundary polygon to the polygon mask layer for use in building the model boundary. The command to "build regular quad grids" was used to specify an appropriate base grid cell size (200m was used for New York City), setting model depths to be stored at the edges of each cell in the grid (Lippert, 2010; Sehili et al., 2014). This base grid resolution was selected such that the primary channel of the Hudson River would have on average, approximately 7-8 base grid cells across the Hudson River, 3-4 across most parts of the East River, and 1-2 across the narrow straits of the Harlem River for proper calculation of water volume transport into and out of the system. Once the regular quadrilateral grid cells were built, the topography, bathymetry, and building ASCII DEMs were imported into the grid editor. Boundary polygons were subsequently generated for the grid using the previously imported polygon in the polygon mask layer. To complete creation of the land boundary, the system editor was used to edit the boundary markers to set the grid boundary marking the edges completely outside of all mask polygons, and then manually unselecting water boundaries along the south, west, and east as open boundaries, and setting the north boundary along the Hudson River as a flux boundary condition.

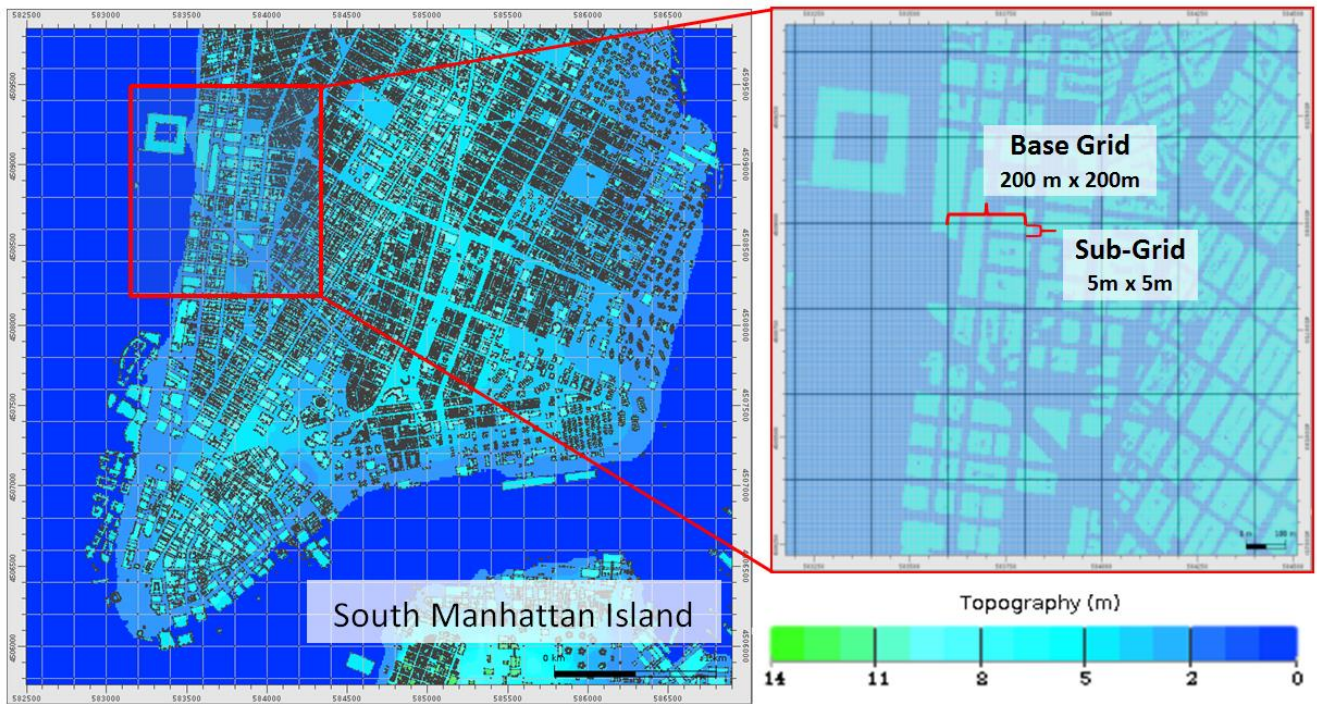

**Figure 2.** Representation of the square sub-grid used for modeling Hurricane Sandy in New York City on the southern tip of Manhattan Island. The grid includes a uniform 200m resolution square computational grid with a nested 5m resolution sub-grid arranged in a 40x40 configuration. Lidar-derived topography data are directly imported into the square sub-grid elements to effectively resolve buildings and streets. Coordinates are in UTM zone 18N.

The bathymetry layer was then merged with the topography layer, with the shoreline forming the seam between the two layers to fill in a complete DEM with topography and bathymetry elevations. Building polygons with the height attribute were added to the bare-earth DEM with buildings as solid 2.5D structures within the DEM At building locations, the height of the DEM was the building height. Finally, the sub-grid was generated with the specification of 40 divisions along each 200m×200m base grid cell edge to produce a nested 5m resolution sub-grid. The combined 5m resolution DEM including lidar-derived topographic heights, detailed bathymetric depths, and buildings, was subsequently translated at its reprojected native resolution onto the 5m resolution model sub-grid. The combined topography, bathymetry, and buildings layer was saved as an .xyz point file with 5m spacing, and the sub-grid mesh was saved as a grid input file for use in modeling inundation caused by Hurricane Sandy in New York City.

To increase the accessibility of our model results to other scientists, policy-makers, and the general public, all geotiffs are converted to geo-referenced images for use with visualization in Google Earth and other preeminent online platforms. Examples of sub-grid model predicted maximum inundation extents are available in three prominent online formats:

Google Earth:
http://web.vims.edu/physical/3DECM/SandyNY/SandyNYMaximums.kmz
Google Maps:
http://web.vims.edu/physical/3DECM/SandyNY/googlemaps.html
Open Layers:
http://web.vims.edu/physical/3DECM/SandyNY/openlayers.html.

Succinctly stated, post-processing procedures rasterize the sub-grid model's coarse computational grid data, combine them with the lidar-derived topography, bathymetry, and buildings stored in the sub-grid, and convert them into usable GIS and Google Earth spatial formats, where the utility of the model predictions may be capitalized upon for statistical spatial comparison and conveniently published in accessible places and formats.

*3.2 Fluxes and volume transport in sub-grid model*

In an urban environment, precise representation of infrastructural features is especially critical when estimating cross-section area. Efficiently resolving infrastructural features is important, as each one uniquely affects the fluid flux through each grid cell side, which ultimately determines the water depth and extent of flooding via distribution of water volume within each grid cell. In this way, sub-grid technology effectively permits velocity to be determined efficiently at the sub-grid scale by sub-dividing computational grid edges (represented as outside bold black lines in Figure 3) in an approach similar to mean value theorem to better approximate cross-section area. In the case of Hurricane Sandy described in this paper, the storm surge flooded into New York City, encountering complex developed land surfaces including bridges and a diverse assortment of buildings, as depicted in Figure 3A-D. These prominent features have a critical impact on storm surge extents, and hydrodynamic modelers are posed with the challenge of either resolving each unique feature, which includes millions of structures in New York City alone, or calculating a porosity function to limit the flood of water allowed to flow through each land grid cell. The first approach, is difficult, because even provided with the abundant computing resources available today, it is still insufficient to model all complex topographic features at the individual building scale or at street-level resolution using the conventional hydrodynamic modeling approach. If we modeled the 4,496,833 sub-grid cells of the New York City domain using 5m resolution with the conventional hydrodynamic modeling method, it would take so long to complete, that any potential forecast would be irrelevant. However, by integrating lidar topographic heights and bathymetric water depths into a high resolution DEM with urban infrastructure, the sub-grid method can more efficiently calculate fluxes through neighboring cell sides within a coarse grid model to further improve model accuracy (Casulli, 2009; Loftis et al., 2013, 2014; Wang et al., 2014; Loftis, 2014).

Given that a flux is the rate of volume flow through an area, the flux times cross-section area is a measurement of the rate of volume flow. Using the sub-grid approach, cross-section area is not simply calculated using one average value for the entire computational grid edge, as in the conventional modeling approach, but is sub-divided into multiple sections to estimate cross-sectional area using the divisions specified in the sub-grid as shown in equation (1):

$$\frac{V_i\left(\eta_i^{n+1}\right)-V_i\left(\eta_i^n\right)}{\Delta t} = -\sum_{l=1}^{S_i} s_{i,\ell} A_j\left(\eta_{j(i,\ell)}^n\right) u_{j(i,\ell)}^{n+\theta} \qquad (1)$$

where indices $i = 1, \ldots,$ # of grid cells, and $j = 1, \ldots,$ # of grid sides, $\Delta t$ is the time step, and $s_{(i,l)}$ is a sign function associated with the orientation of normal velocity on the $j^{th}$ side. Specifically, for the $i^{th}$ computational base grid cell, the volume of the water column at different times $(n, n+1)$ for specified water levels, $\eta$, are defined as $V_i(\eta_i^n)$ and $V_i(\eta_i^{n+1})$, such that the flux is characterized by the left hand side of (1), represented by the blue arrows in Figure 3A-D.

Velocity at each side, $s_i$, of every grid cell is $u_{j(i,l)}^{n+\theta}$, with $\theta$ representing a semi-implicit coefficient. The cross-section area, $A_j(\eta_{j(i,l)}^n)$, on each grid side for a specified water level, $\eta_{j(i,l)}^n$, varies with local topography and bathymetry, and in the presence of urban infrastructure, such as buildings and bridges, can be perceived as a porosity function, limiting fluid flux through each grid cell side. However, no porosity function is necessary using the sub-grid approach, as it is inherently accounted for if impediments to fluid flow are resolved in the DEM used to produce the sub-grid. Essentially, in the sub-grid model, the total flux into a specific grid cell multiplied by each cross-section area is equal to the rate of total volume change in the cell. Thus, the fluxes through each grid cell side and the water volume in each cell may be more accurately approximated, leading to more accurate non-linear volume transport calculations (Casulli, 2009; Casulli and Stelling, 2011).

As depicted in the illustration in Figure 3A, if buildings are not considered in the model simulation, then a porosity function would allow 100% of fluid flow to pass through all edges of the grid cell. However, when buildings are included in Figure 3B, the left edge is blocked by buildings occupying 2 of the 6 sub-grid edges. Without the sub-grid approach, a porosity function should limit fluid flux through this side into the cell by 66.6%. Also, 3 of the 6 edges on the right side are obstructed by buildings, causing making an appropriate porosity function limit flux out of the cell to 50%. In a likewise fashion, the top side would be limited to 66.6% flow with buildings obstructing 2 of the 6 sub-grid edges, and the bottom side would be limited to 33.3% flux having only 2 unobstructed sub-edges. This example using one sub grid cell demonstrates the complexity of inundation modeling in an urban environment, as this function would have to be uniquely calibrated to characterize each of the 23,559 computational grid sides without efficiently resolving the buildings via the sub-grid approach. No porosity function is necessary using the sub-grid approach, as it is inherently accounted for if impediments to fluid flow are resolved in the DEM used to produce the sub-grid.

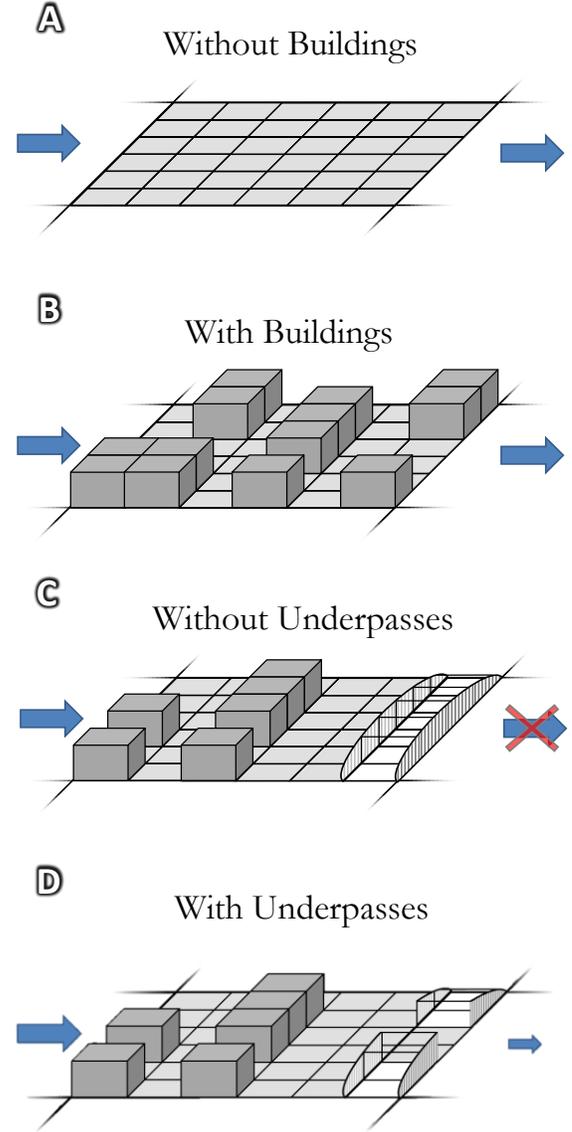

**Figure 3A-D.** Examples of discrepancies between FEMA maximum inundation extents and sub-grid model-predicted inundation due to the presence of buildings and improper representation of bridge underpasses blocking fluid movement included in the model's lidar-derived DEM, before (A and C) and after (B and D) rectification in the model sub-grid.

Roadway overpasses are another considerable source of potential error due to the primary method for topographic data collection being final-return lidar measurements for high resolution DEMs. Figure 3C illustrates a highway overpass effectively blocking the entire right side of a computational grid cell, occupying each of the sub-edges, which typically results in the model over-predicting flooding on the side closest to the water, and under-predicting inundation on the other side. However, using the GIS roadway extraction methodology outlined in this study, the vertically occluded bridge underpasses may be "carved out" using appropriate elevation values of neighboring roadway topography measurements to allow for a free path through the grid cell side, and properly characterize fluid flux through otherwise blocked grid cells for a more accurate spatial prediction of inundation(Figure 3D).

High-resolution hydrodynamic models are needed to appropriately consider the impact of these local features into the prediction of maximum storm surge extents. In Casulli's (2009) paper, the sub-grid model's solution algorithm is referred to as a mildly nonlinear system for the free surface wherein the formulation for finite volume leads to a mildly nonlinear system for finite volume with respect to the free surface elevation. This nonlinear solver operates on base grid cell sides, and is non-linear, because as volume increases, the slopes of the river banks are not uniform (Aldrighetti and Zanolli, 2005). Since the 'container' holding the fluid is a complex shape, and therefore not prism-shaped idealized flat walls perpendicular to a flat river bottom, the fluid volume increases and decreases nonlinearly with the rise and fall of the free surface of the water (Casulli and Zanolli, 2012). Given the anomalous rise in the free surface of 3.5m observed at The Battery, NY, Hurricane Sandy's storm surge induces a nonlinear increase in volume transport as the flood waters are not constrained by the riverbanks and freely flood into the streets of New York City. Since cross section area is not simply calculated using one average value for the entire base grid edge, as in the conventional modeling approach, but is sub-divided into multiple sections to estimate cross-sectional area using the divisions specified in the sub-grid; the fluxes through each grid cell side and the water volume in each cell may be more accurately approximated, leading to more accurate non-linear volume transport calculations (Casulli, 2009; Casulli and Stelling, 2011).

*3.3 Bottom friction*

The model's specified bottom friction for overland flow is verified by the results of a small scale laboratory experiment to ascertain flow resistance to storm surge induced inundation in the presence of buildings (Wang, 1983). These results were then scaled to average building spacing using dimensional analysis and through proper scaling of building disposition parameters from the laboratory experiment to average building spacing within the city blocks in each New York City borough. Separate drag coefficient equations, may be calculated for each of three building classifications in New York City: high rise, medium rise, and residential.

Application of these equations requires knowledge of building density and building classification. This information may be calculated for New York City utilizing GIS tools on the building layer embedded within the sub-grid model. Considering that most of the buildings in New York City are aligned in configuration to maximize transportation efficiency, and that virtually all of the buildings along the water or within the flood risk area fall into the classification of high-rise buildings in the urban metropolis; the following method was utilized to calculate building density for each of the boroughs in New York City in the interest of applying a spatially-varying over land friction coefficient, $C_{Db}$. This $C_{Db}$ will be specified in the sub-grid model's 2-D friction formulation using Manning's formula with a spatially-varying bottom roughness coefficient, $n$, (Table 1) by way of a conventional similarity solution (Wang and Christensen, 1986).

Using GIS tools, the building areas were retrieved from the vector dataset. Interior terrestrial base grid cells (not including grid cells containing portions of the river) were selected as a representative sample of building density within each 200m x 200m base grid square for each of the boroughs within the sub-grid domain. Table 1 includes spatial analysis results for building density and analogous measures of $C_{Db}$ with translated values for Manning n using Wang's suggested $C_{Db}$ for high rise buildings in an aligned configuration (1983). Overland values for Manning n are spatially varying by building density within each New York City borough according to Table 1. Building density ratios calculated from the 200x200m base grid cells for each borough were converted to 1x1m scale to yield $C_{Db}$ values ranging from 0.2365 in Staten Island to 0.2813 in Manhattan. These values translate to Manning n values for a range from 0.0896 in Staten Island to 0.0978 in Manhattan (Table 1). Standard Manning n values of 0.020 in the Hudson River, and 0.030 in the East River and Harlem River were used to represent bottom drag within the New York Harbor. Both of these values are reasonably close to the average Manning n value of 0.025 for relatively straight river channels (Henderson, 1966).

Provided the use of high-resolution lidar-derived topography data and extremely accurate vector building data, streets between buildings may be sufficiently resolved within the model sub-grid to intrinsically account for the form drag posed by the storm surge flow around building obstacles. The arrangement and configuration of buildings along with the disposition between rows of buildings along the water's edge vary greatly by shape and size (Wang, 1983). Each of these building shapes would need to be uniquely accounted for in the model's friction specification if their shape is not resolved within the hydrodynamic model grid. This is a task that is either impossible or highly impractical due to computational demand when using the conventional modeling approach. While the inland metropolitan area surrounding New York City is generally structured in a block system to maximize utility for the urban population, buildings adjacent to the water's edge often have unique shapes, being designed to maximize the number of rooms with a view of the adjacent body of water. Each of the buildings varies by shape and dimension, and thus has their own unique form drag. This unique form drag is in addition to the friction posed by the ground surface, both of which must be accounted for in the model's friction parameterization if the model grid does not sufficiently resolve buildings. Thus, the sub-grid model effectively resolves the streets using high-resolution topography to utilize a more universal friction specification.

**Table 1.** Spatial analysis results for building density and average diagonal building disposition within each New York City borough.

| # | Borough | 200x200m Cells | Total Cell Area (m²) | Building Area (m²) | Coverage Ratio ($M$) | $C_{Db}$ | $n$ |
|---|---|---|---|---|---|---|---|
| 1 | Manhattan | 1198 | 47,920,000.00 | 19,410,903.63 | 0.4051 | 0.2813 | 0.0978 |
| 2 | Brooklyn | 1271 | 50,840,000.00 | 16,159,370.29 | 0.3178 | 0.2595 | 0.0938 |
| 3 | Queens | 1535 | 61,400,000.00 | 18,432,395.83 | 0.3002 | 0.2551 | 0.0931 |
| 4 | Bronx | 1101 | 44,040,000.00 | 10,685,412.81 | 0.2426 | 0.2407 | 0.0905 |
| 5 | Staten Island | 402 | 16,080,000.00 | 3,635,814.19 | 0.2261 | 0.2365 | 0.0896 |

*3.4 Model setup and simulation*

Atmospheric data for the observation simulation of 2012 Hurricane Sandy in the New York Harbor region were collected in units of m/s from NOAA atmospheric observation data at Bergen Point, New York (NOAA Station # 8519483). Atmospheric observations were subsequently pre-processed and prepared as uniform wind and pressure inputs throughout the small-scale domain. U and V velocities were extracted and wind fields were interpolated to 6-minute time steps, commencing on October 01, 2012, at 00:00 GMT, and ending November 04, 2012, at 00:00 GMT. Atmospheric pressure was converted from mbars to m²/s², and prescribed as a uniform atmospheric pressure input throughout the domain in similar fashion to the prescribed wind inputs.

Tides are forced via three open boundaries: one to the south, one in the west, and one in the east. The southern open boundary in the sub-grid domain is located at the mouth of the New York Bay into the Raritan Bay leading to the Atlantic Ocean. The open boundary to the west is where the Kill van Kull connects the Newark Bay to New York Bay. The third open boundary lies to the east and connects the East River to the Long Island Sound. The southern boundary is forced using observation data from USGS Rockaway Inlet (Station #1311875), the west boundary uses NOAA Bergen Point (Station #8519483), and the east boundary is forced using water level data from NOAA Kings Point (Station #8516945) shown in Figure 1. The forcing data from Rockaway Inlet was converted from NGVD29 to and delayed by 12 minutes to account for its distance from the southern boundary of the grid at Coney Island, south of the Verrazano Narrows.

Hourly freshwater flows for the Hudson River were obtained from the USGS and specified as a flux boundary condition. These data were applied uniformly as a forcing along the sides of 9 elements along the northern boundary of the model domain near Wappingers Falls (Station #01372500). The freshwater flow input has been delayed by 30 minutes to account for the considerable distance from the station to the northern edge of the sub-grid domain. Additional information regarding model setup and calibration via time series comparisons and USGS rapid-deployment gauges are described in further detail in Loftis, 2014, and Wang et al., 2014 (Figure 4A-C). Average simulation time was ≈70min to complete each 35 day run.

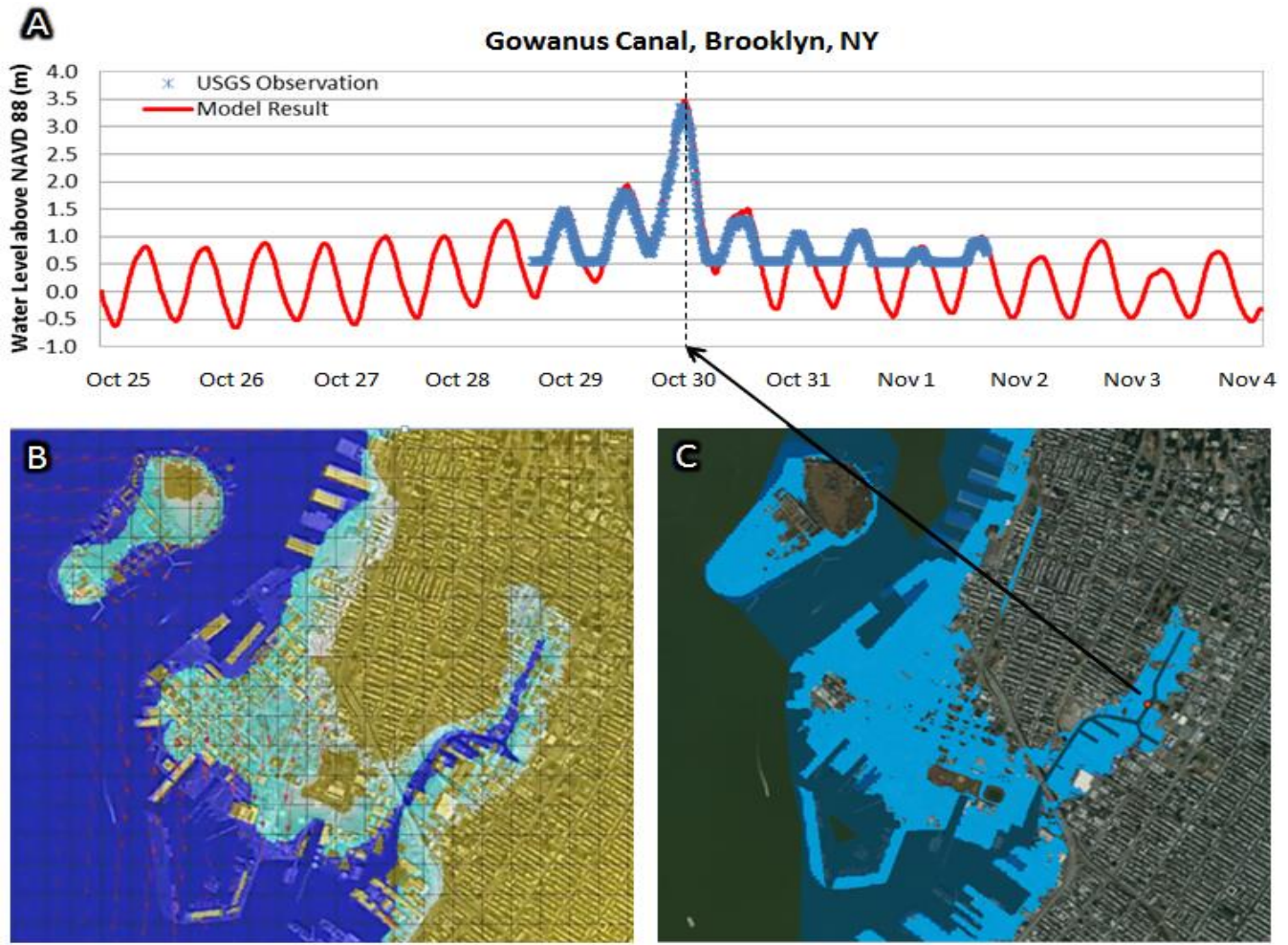

**Figure 4A-C.** (A) Simulated storm tide time series (red) comparison with USGS-collected rapid deployment gauge (blue) in the Gowanus Canal, Brooklyn, NY (shown as a red dot in 4C). (B) The simulated maximum horizontal inundation extent rendered by the sub-grid model. (C) Maximum inundation extent estimated by FEMA using the USGS Hurricane Sandy storm tide mapper. Additional time series comparisons featured in Loftis, 2014, and Wang et al., 2014.

## 4. Observation data and post-processing model results

Spatial observation data retrieved in the aftermath of Hurricane Sandy includes 73 USGS-collected non-wave-affected high water mark measurements within the New York Harbor sub-grid model domain which will be used to conduct a point-to point comparison with model results (McCallum et al., 2013). The USGS surveyed 653 independent high water mark locations in the aftermath of Hurricane Sandy ranging from Virginia to Massachusetts. These marks, noted as water stains or debris markings such as dirt or seed lines were used as a benchmark for model comparison considering the maximum extent of inundation. The measurements were typically made along sides of buildings or lamp posts, or via debris lines washed ashore near the ground, and were surveyed relative to NAVD88, with a plurality of measurements collected in New York and New Jersey where the impacts of the storm were the most heavily pronounced. Within the extent of the sub-grid model domain, there were 62 non-wave affected high water mark observation sites in New York City, and 11 non-wave affected marks in the State of New Jersey for comparison. A high water mark was considered to be an independent measurement location if separated by more than 1,000 feet from neighboring high water marks (McCallum et al., 2013).

Additionally, 80 FEMA-reported inundated school locations indicating water level thickness at specific sites throughout the sub-grid domain will be the subject of a point to point comparison. The FEMA inundated schools data set is a homeland infrastructure geospatial data inventory of 295 schools damaged to various degrees via flooding during Hurricane Sandy, and assembled by the National Geospatial-Intelligence Agency in partnership with the Department of Homeland Security in 2012 (FEMA MOTF, 2013). Data for public and private schools in New York were provided by the New York State Department of Education in New York City only. New Jersey public and private schools were furnished via the New Jersey Department of Education with the data being available as a GIS shape file.

Finally, distance and area comparisons with model results will be made using a maximum extent of inundation map produced by FEMA's Modeling Task Force (FEMA MOTF, 2013). The maximum inundation extent map product is created from storm surge sensor data, and field-verified high water mark data collected by the USGS post-Hurricane Sandy (McCallum et al., 2013). These data products are subsequently utilized to interpolate a water surface elevation, then subtracted from the best available DEM to create an inundation grid and surge boundary utilizing a GIS bathtub model for each state substantially affected by the storm. In this sub-grid model comparison, the final released datasets for the 3m-New Jersey and 1m-New York City products released on February 14, 2013, were utilized for spatial comparison with model results (FEMA MOTF, 2013).

Creation of an inundation map for Hurricane Sandy in the New York Harbor requires substantial interoperability through GIS-compatible formats, including conversion from unstructured grid model element data for water elevations. Upon conclusion of a model simulation, combined water elevation and velocity results are passed to a python script provided with a copy of the model grid and Geospatial Data Abstraction Libraries (GDAL) for translation of elevations and velocities to a set of geotiffs (GDAL, 2014). One geotiff is produced for each specified model output time step, with resolutions at the scale of the base grid for water elevations and velocities at cell center points throughout the domain. The results of this operation were passed to a Linux shell script, relating water level elevations from the coarse computational grid to the topography and bathymetry data of the sub-grid using the open source GIS software, GRASS, via the r.mapcalc() command, resulting in two new sets of geotiff rasters: 1) water elevation data (meters above NAVD88), and 2) water thickness data (m above local elevation), both at the resolution of the sub-grid at 5x5m pixels (GRASS Development Team, 2012). Subsequently, the script surveyed each sub-grid pixel of the output rasters across all time steps to export the maximum recorded value for inundation into an 'elevmax.tif', for maximum predicted water elevation, and 'thickmax.tif' for maximum predicted inundation thickness. The 'elevmax.tif' product was used to assess the model maximum water elevation extent against USGS high water mark data (also measured relative to NAVD88), and the 'thickmax.tif' geotiff was utilized for comparison with FEMA's inundated schools dataset (measured relative to the local ground surface).

Additionally, a copy of the layer was converted from a geotiff raster to a polyline shapefile, extracting and saving the outermost maximum inundation extent contour as 'thickmax_line' for use in distance comparisons. The 'thickmax_line' shapefile and 'thickmax.tif' geotiff were subsequently utilized in statistical distance and area comparisons against an inundation map distributed by the FEMA Modeling Task Force. (FEMA MOTF, 2013). The distance measurement methodology utilizes 3m-New Jersey and 1m-New York City clipped polygons produced by FEMA as a mask for inundated areas. In the distance assessment, the outermost inundation extents were interpreted to be the maximum extent of inundation, so as to ignore impediments to flow such as buildings. The FEMA maximum inundation extent line was converted from a line to a series of points with 5m regular point spacing (similar to the sub-grid resolution) along the line via the construct points toolset within the ArcGIS 10.1 editor. Subsequently, the near/distance calculation feature utilized the standard distance formula to export a table containing shortest distance calculations to the model predicted maximum inundation line for each of the nearly 100,000 5m-spaced points along the FEMA maximum inundation line.

In preparation for performing an exhaustive area comparison between the FEMA maximum inundation data and the model predicted maximum inundation map, both polygon layers were collected and clipped using the shoreline and building layer to remove over-water areas and buildings from both datasets, such that only flooded land area is assessed in the comparison. The resulting polygon layers were converted to 5m resolution rasters, and subsequently mosaicked with a raster of the entire region, assigning a default data value of '3' for non-inundated sub-grid cell pixels, and a value of '2' for inundated areas in the sub-grid model raster, and '1' for areas inundated using FEMA's maximum extents. Notably, without this critical step, the following raster math 'mapcalc' function will only assess the difference of regions shared by both the FEMA inundation raster and the sub-grid model's spatial maximum, consequently ignoring the differences (under-predicting and over-predicting regions) between the two rasters due to no-data values. The sub-grid maximum extent raster was subtracted from the FEMA maximum extent raster, yielding four field values in the resulting difference map: 2-1=1 (match), 2-3=-1 (over-predict), 3-1=2 (under-predict), and 3-3=0 (no flooding).

Finally, the resulting difference raster is converted to polygons, without smoothing or otherwise simplifying the polygons, to make use of the area calculation toolset. The total areas are calculated for each polygon and aggregated in a table to provide relevant statistics for total area (m²) and percent area (%) of matching/intersecting agreement between the FEMA observation data and the model prediction along with errors where the model over-predicted and under-predicted the recorded data. After assessing the total difference areas, the New York Harbor region was separated by river system to address areas analogous to the distance comparison and focus on locations where the model performed well and investigate areas where it did not.

## 5. Geospatial comparison of results

Verification of the spatial extent and depth of flood waters within the New York Harbor sub-grid domain was assessed via comparison of model-predicted results with a variety of verified-field measurements from various agencies. First, 73 USGS-collected non-wave-affected high water mark measurements within the New York Harbor were collected for comparison with water level elevation above NAVD88 in meters. Second, 80 FEMA-collected inundated school locations where flood waters left visible moisture marks indicating water level thickness at specific sites throughout the sub-grid domain, and were also used as a point-to-point comparison method (Included in Appendix). Third, a variety of distance and area coverage calculations are utilized to compare model results with FEMA's maximum extent of inundation map, which was based upon interpolation of the USGS's high water marks and the best available elevation data.

*5.1 Point-to-point comparisons with USGS-recorded high water marks*

The model comparisons for high water marks were separated by state and by county by the USGS, with counties in the study area noted in Figure 1. However, statistics were not computed by county, since the gerrymandered municipal boundaries have minimal impact on the extent of inundation from a hydrodynamic standpoint. Most of the high water marks were measured on Manhattan Island (or New York County, abbreviated as NEW in Appendix A), with a range in difference between the observed high water mark and maximum water level height reported by the model from 0.0168 to 0.2639m. Most of the other water marks were collected in Queens (abbreviated as QUE) ranging from 0.0710 to 0.2970m in difference, or in Brooklyn (or Kings County, abbreviated as KIN) ranging in difference from 0.0258 to 0.2788m. The remaining two boroughs surveyed had 3 measurements from the Bronx (abbreviated as BRO) ranging from 0.1187 to 0.2000m, and from 2 measurements from Staten Island (or Richmond County, abbreviated as RIC) ranging from 0.2271 to 0.2971m, with larger differences than the other areas likely due to the proximally close position to the mouth of the New York Bay with some small wave effect noted at these stations (Appendix A). A few high water marks in this area of Staten Island and its analogous position across the Bay on Coney Island were noted by in the USGS report to be heavily affected by waves. These high water marks were not included in the spatial statistical comparison due to the model not addressing wind-wave interaction, and due to the relative uncertainty of water mark measurements accurately representing the average flood height for prolonged periods in areas frequently buffeted by waves.

In the state of New Jersey, a majority of the 12 high water marks were recorded in Hudson County (abbreviated as HUD). The 10 high water marks had a large range in difference from 0.1261 to 0.5290m. The differences in the remaining 2 measurements collected from Bergen County (abbreviated as BER) within the study area were also large, with values of 0.5406 and 0.5577m. The large differences were anticipated due to the lack of freely available building height data for the New Jersey side of the Hudson River being represented in the model's DEM. Subsequently, without the building presence in the grid, the modeled flooding extent was greatly exaggerated in places where high building densities would have buffered or impeded fluid flow, such as Jersey City, Hoboken, and areas of Bayonne (Appendix A). Aggregated statistics for New York presented in Table 2 suggest a very favorable comparison with a mean difference of -0.0004m indicating no strong bias towards over-prediction or under-prediction of high water marks by the model. The mean difference of 0.2150m reported for New Jersey suggests that the model tended to over-predict recorded high water marks by 21.5cm on average. The absolute mean difference was 0.112m for New York with a value 3x greater being calculated for New Jersey at 0.364m. The smaller ranges described previously in the high water marks for the different boroughs of New York City logically led to a relatively small standard deviation of 0.085m and an RMSE of 0.120m when compared to standard deviation in New Jersey at 0.256m and an RMSE of 0.347m. The difference of 0.227m is a significant indication that the inclusion of buildings in the model DEM is critical to urban inundation modeling.

*5.2 Point-to-point comparisons with FEMA inundated schools dataset*

Within the study area of New York City, most of the inundated schools were located in Manhattan and Brooklyn. The worst flooding was observed at schools neighboring Coney Island Creek along the more coastal areas of New York City (Appendix B). Statistical measures for New York City are reasonably favorable with a mean difference of 0.0332m, implying no leaning towards over-prediction or under-prediction of inundated schools by the model (Table 2). The mean difference of 0.3483m reported for New Jersey suggests that the model tended to over-predict recorded high water marks by 34.8cm on average. The absolute mean difference was 0.2769m for New York, compared to 0.4227m calculated for New Jersey. The standard deviations in the two data sets were about equal with 0.3304m for schools in New York and 0.3328m in the model comparison against flood heights at schools in New Jersey (Table 2).

The impact of waves affected FEMA's inundated schools dataset due to its relation to the USGS high water marks while the sub-grid model results do not. Thus, regions with higher wave influence may have exaggerated water levels in the FEMA dataset, around the Southern New York Bay and Staten Island, extending the range of the calculated differences between the sub-grid model and the inundated schools for New York (Table 2). The RMSE for the 60 schools in New York City within the sub-grid domain was 0.3293m. Upon comparison with the RMSE of 0.4760m for the 20 schools in New Jersey, the point-to-point evaluation with the New Jersey schools led to 0.1467m more RMSE. As with the other point-to-point comparisons using the USGS high water marks, the RMSE difference of 0.1467m more error in New Jersey is likely attributed to the lack of freely available building data in the model's DEM.

**Table 2**. Statistical metrics for mean difference, absolute mean difference, standard deviation, and root-mean-squared error (*RMSE*) for sub-grid model comparisons with 73 USGS-observed non-wave affected high water marks, and 80 depths at inundated schools reported by FEMA within regions of New York and New Jersey within the study area.

| Data Location | USGS High Water Marks | | | |
|---|---|---|---|---|
| | Mean Diff. | \|Mean Diff.\| | Std. dev. | *RMSE* |
| New York HWMs | -0.0004 | 0.112 | 0.085 | 0.120 |
| New Jersey HWMs | 0.215 | 0.364 | 0.256 | 0.347 |
| | FEMA Inundated Schools | | | |
| | Mean Diff. | \|Mean Diff.\| | Std. dev. | *RMSE* |
| New York Schools | 0.033 | 0.277 | 0.330 | 0.329 |
| New Jersey Schools | 0.348 | 0.423 | 0.333 | 0.476 |

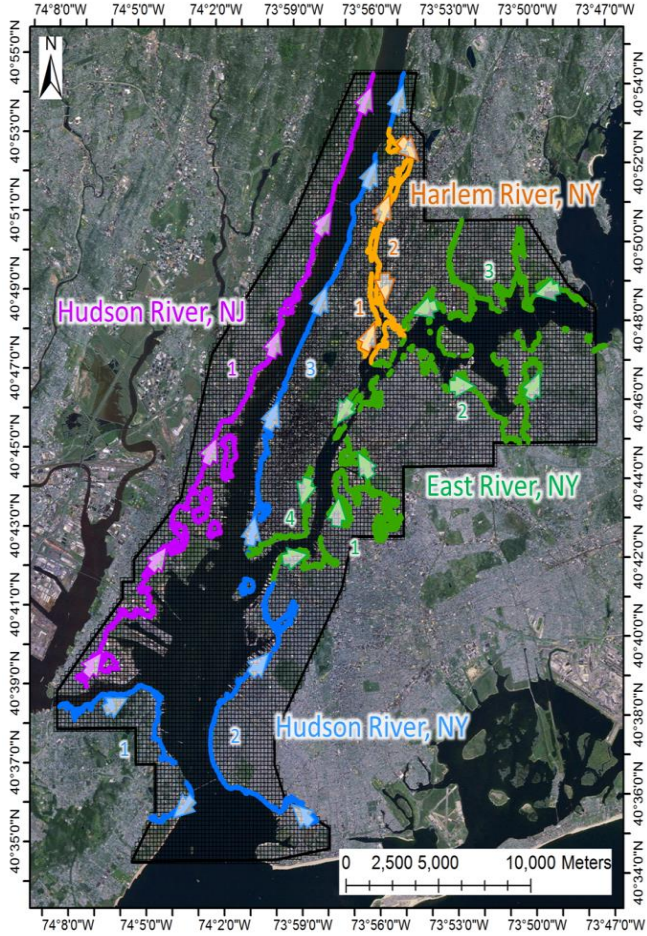

**Figure 5.** Distance measurement map displaying the observed maximum extent of inundation reported by FEMA, separated by color into four regions by river system and state. Numbers and arrows illustrate the direction and order of distance measurements following along each region corresponding with Figure 6A-D.

*5.3 Spatial comparison with FEMA maximum extent of inundation map*

Spatial area comparison with FEMA's maximum extent of inundation map was two-fold. This spatial flood coverage map was based upon interpolation of the USGS's field measurements including high water marks and rapid deployment gauges compared in the previous sections and the best available digital elevation data. The field-verified high water mark measurements collected in the aftermath of Hurricane Sandy were utilized to construct an interpolated GIS layer of water surface heights, which was subsequently subtracted from the best available DEM to create a water level thickness layer and a 0m contour for the maximum extent of inundation. These products are comprised of an inundation grid at 1m resolution for New York City and 3m resolution for New Jersey, along with a clipped surge boundary (FEMA MOTF, 2013). The database and GIS products produced by the USGS and FEMA were enormously valuable as a standard for spatial comparison with the sub-grid model results. These data were collected to calculate distances between the model's predicted maximum flood extent and FEMA's reported maximums (Table 3), and to compute inundation percent area match statistics for additional spatial verification of the model (Table 4).

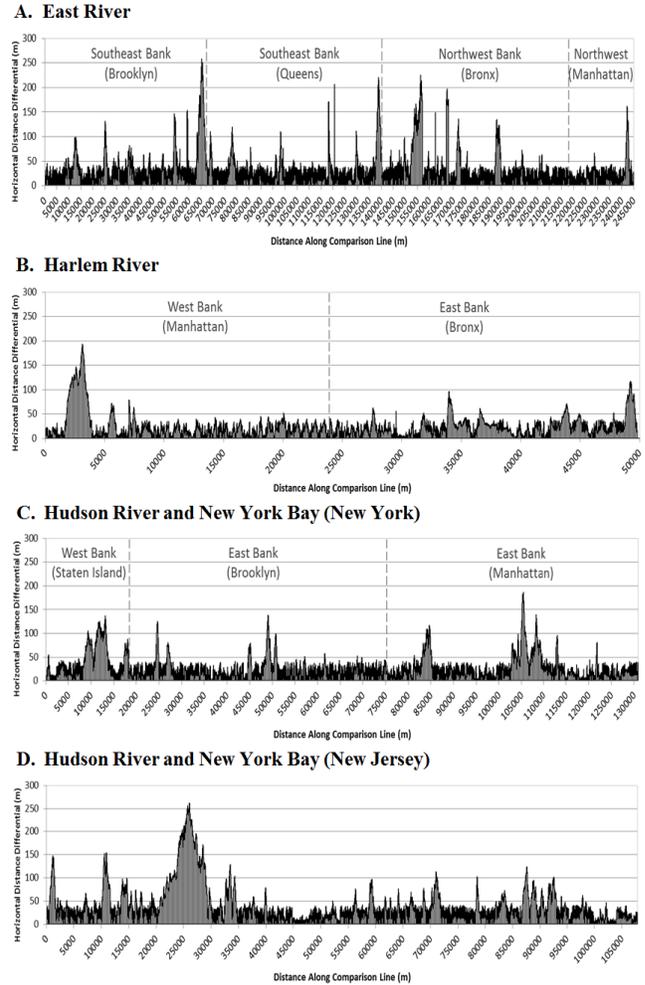

**Figure 6A-D.** Plotted absolute distances to the model's predicted maximum extent of inundation in reference to the observed maximum extent of inundation line reported by FEMA. Distance measurements are separated into four regions by river system, including: East River (A), Harlem River (B), Hudson River on the New York side (C), and along the New Jersey coast (D).

**Table 3.** Distance difference table featuring rows separated by region, and columns for metrics of absolute mean distance and standard deviation for initial sub-grid simulations not accounting for highway underpasses on the left and with underpasses on the right. All units are in meters.

| Survey Region | # of Points | No Underpasses |  | With Underpasses |  |
|---|---|---|---|---|---|
|  |  | \|Mean Dist.\| | Std. dev. | \|Mean Dist.\| | Std. dev. |
| East River, NY | 47,283 | 46.78 | 58.31 | 39.68 | 36.11 |
| Harlem River, NY | 9,673 | 44.22 | 56.70 | 34.02 | 32.32 |
| Hudson River, NY | 21,492 | 28.88 | 27.02 | 27.01 | 23.22 |
| *All New York* | 78,448 | 39.96 | 47.34 | 33.57 | 30.55 |
| Hudson River, NJ | 16,396 | 36.90 | 30.38 | 27.80 | 22.95 |
| *All New Jersey* | 16,396 | 36.90 | 30.38 | 27.80 | 22.95 |
| *All Hudson River* | 37,888 | 32.89 | 28.70 | 27.40 | 23.09 |
| *All Study Area* | 94,844 | 38.43 | 38.86 | 32.13 | 28.65 |

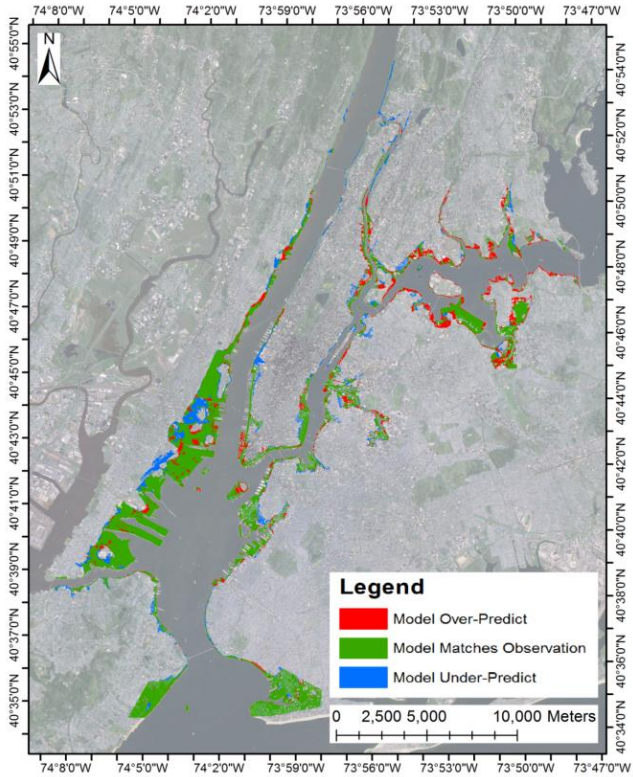

**Figure 7.** Area comparison with FEMA maximum extent of inundation map in the New York Harbor region during 2012 Hurricane Sandy superposed with satellite imagery. Shaded areas are 5m² sub-grid cells highlighted according to whether the sub-grid model over-predicted (red), matched (green), or under-predicted (blue) the spatial extent of inundation coverage reported by FEMA.

*5.3.1 Distance differential assessment*

Maximum spatial extent of inundation is an especially critical attribute to address in assessing flooding risk. The initial distance comparison utilizing points along the FEMA-clipped maximum extent of inundation line revealed a relatively favorable distance differential with the model-predicted maximum inundation across the sub-grid domain with an absolute mean distance difference of 38.43m (Table 3). Upon evaluation of maximum inundation distance by river system, the absolute mean distance indicated minimal difference along the Hudson River and New York Bay region with a 28.876m difference along the New York City Bank, and 36.9m along the lower elevation New Jersey bank. The Hudson River was divided by state instead of west/east bank due to the lack of freely available building data for the New Jersey side for representation in the sub-grid model DEM (Figure 5). The observable difference of 8.024m between the New York bank of the Hudson River (buildings included) and the New Jersey side (bereft buildings) is an indication of the importance of resolving building infrastructure in the model sub-grid for accurate high-resolution inundation prediction.

The average measured distances from the FEMA-reported maximum flood extent points to the model-predicted inundation along the New York bank of the Harlem River were 44.222m, with a 46.779m difference recorded along the East River. The horizontal distance differentials cover a range from 0 to 258.6m (Figure 6A-D). Of the four river systems, the East River accounts for a plurality of the point to line distances with 47,283 points out of the total 94,844 points with 5m regular point spacing along the FEMA maximum inundation line. Together, the New York side (21,492 points) and the New Jersey bank (16,396 points) of the Hudson River account for a 32.888m absolute mean distance, the most favorable inundation comparison of the three river systems (Table 3).

*5.3.2 Area difference map evaluation*

The initial spatial comparison shown in Figure 7 resulted in an overall 75.15% spatial match with 11.41% area model over-prediction and 13.44% model under-prediction (Table 4). Area comparisons along the main stem of the Hudson River performed reasonably well with a 78.80% match along the New York river banks, and a slightly lower match of 76.73% match along the New Jersey river banks. Flooded area was higher for the New Jersey side of the Hudson River, as the 76.73% matched inundation area corresponded to 17,539,367m², while the 78.80% match on the New York side of the river represents 13,076,031m² (Table 4). The ratio of under-prediction to over-prediction was slightly less than 2:1 for the Hudson River with the New York bank having 13.76% under-prediction, representing 2,283,797m², and 7.44% over-prediction signifying a representative area of 1,234,304m². The Hudson River banks adjacent to New Jersey observed slightly more error than their New York counterparts with 14.86% under-predicting FEMA's maximum inundation estimates with an area of 2,283,797m², and 8.41% over-prediction representing an area of 1,922,727m².

**Table 4.** Statistical comparison results for inundated areas in the New York Harbor region with and without highway underpasses incorporated into the sub-grid model. Values are presented as surface areas (m²) and (% area coverage) for each of the defined categories: match, model under-predict, and model over-predict compared with the FEMA maximum inundation coverage map.

| Survey Region | No Underpasses Match | (%) | With Underpasses Match | (%) | No Underpasses Under-Predict | (%) | With Underpasses Under-Predict | (%) | No Underpasses Over-Predict | (%) | With Underpasses Over-Predict | (%) |
|---|---|---|---|---|---|---|---|---|---|---|---|---|
| East River, NY | 14,180,524 | 71.81 | 14,713,802 | 74.51 | 2,211,023 | 11.20 | 1,911,532 | 9.68 | 3,357,069 | 17.00 | 3,124,102 | 15.82 |
| Harlem River, NY | 4,457,765 | 70.34 | 4,528,503 | 71.46 | 918,108 | 14.49 | 891,501 | 14.07 | 961,151 | 15.17 | 917,051 | 14.47 |
| Hudson River, NY | 13,076,031 | 78.80 | 13,160,187 | 79.31 | 2,283,797 | 13.76 | 2,213,695 | 13.34 | 1,234,304 | 7.44 | 1,220,202 | 7.35 |
| *All New York* | 31,714,320 | 74.31 | 32,402,492 | 75.92 | 5,412,928 | 12.68 | 5,016,728 | 11.75 | 5,552,524 | 13.01 | 5,261,355 | 12.33 |
| Hudson River, NJ | 17,539,367 | 76.73 | 20,215,043 | 88.43 | 3,397,304 | 14.86 | 1,019,683 | 4.46 | 1,922,727 | 8.41 | 1,623,731 | 7.10 |
| *All New Jersey* | 17,539,367 | 76.73 | 20,215,043 | 88.43 | 3,397,304 | 14.86 | 1,019,683 | 4.46 | 1,922,727 | 8.41 | 1,623,731 | 7.10 |
| *All Hudson River* | 30,615,398 | 77.60 | 33,375,230 | 84.59 | 5,681,101 | 14.40 | 3,233,378 | 8.20 | 3,157,031 | 8.00 | 2,843,933 | 7.21 |
| *All Study Area* | 49,253,687 | 75.15 | 52,617,535 | 80.28 | 8,810,232 | 13.44 | 6,036,411 | 9.21 | 7,475,251 | 11.41 | 6,885,086 | 10.51 |

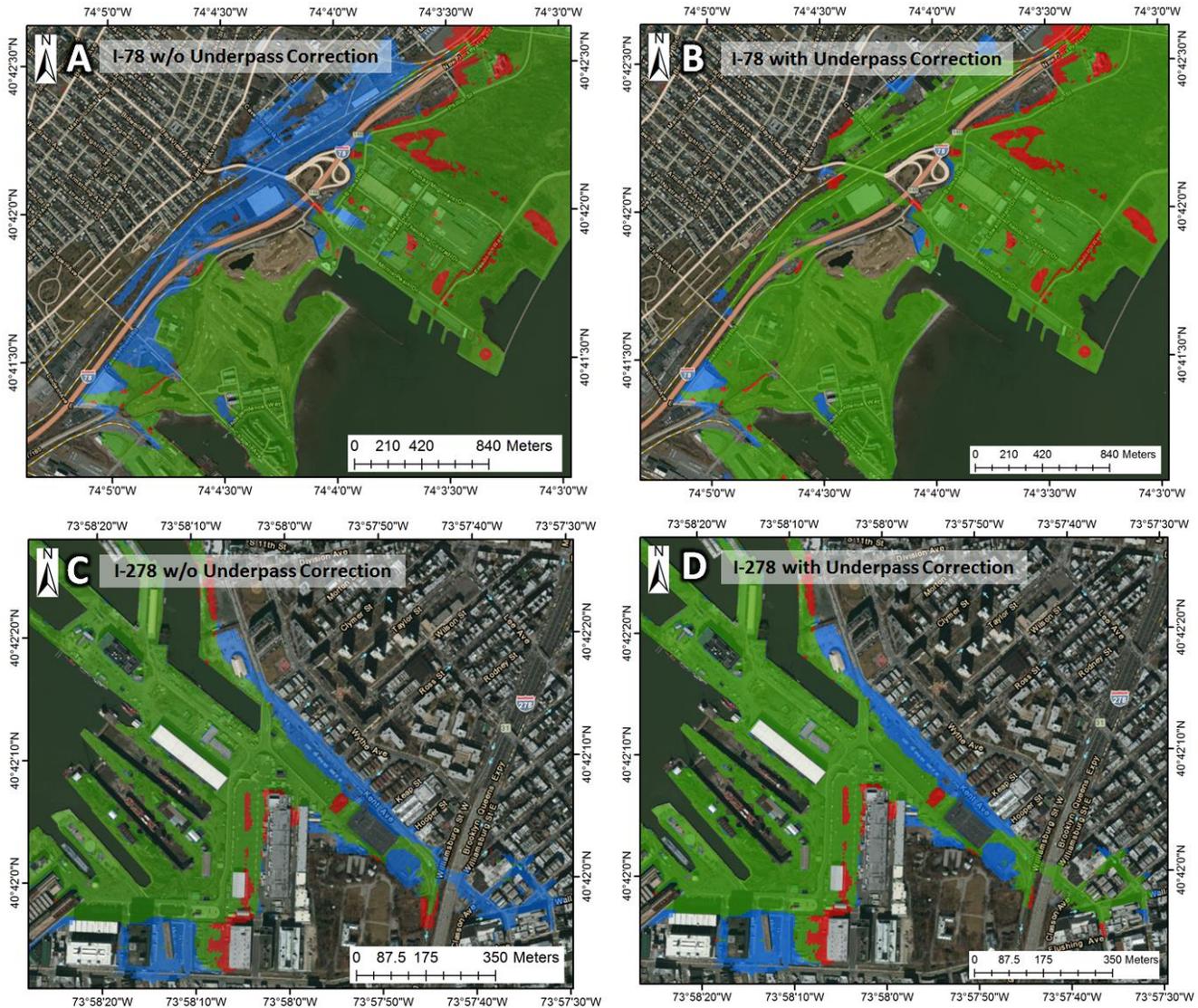

**Figure 8A-D.** Examples of discrepancies between FEMA maximum inundation extents and sub-grid model-predicted inundation due to the presence of roadway infrastructure and overpasses blocking fluid movement included in the model's lidar DEM, before (A and C) and after (B and D) rectification in the sub-grid.

Inundation area comparisons along the East River observed a 71.81% match, and the Harlem River had a 70.34% match between the model and FEMA's maximum inundation map. The under-predicted area was approximately the same as the Hudson at 11.20% (2,211,023m²) and 14.49% (918,108m²), for the East and Harlem Rivers, respectively. However, the over-predicted areas were approximately double those observed in the Hudson River for New York and New Jersey with 17.00% (3,357,069m²) for the East River and 15.17% (961,151m²) for the Harlem River (Table 4). The inundation area of the Harlem River was the smallest due to the smaller and narrower size of the river, and the higher frequency of over-prediction along the East River is attributed to the aforementioned convergence of the two storm surges from the south by Atlantic Coast via the Raritan Bay, and from the east through the Long Island Sound.

Discrepancies between the model predictions and the FEMA flood map are attributed to DEM differences, and possibly the lack of building representation in the FEMA maximum inundation map (Figure 8A-D). Additionally, the implementation of the spatial flooding observation data as a derivative "bathtub model" product of USGS-interpolated high water marks and elevation data without regard for strong water current velocities or estuarine circulation could account for regions with significant discrepancies (Schmid et al., 2014). Such discrepancies can be addressed in both the area and distance spatial comparisons to minimize the impact of DEM incongruities that are outside of control for the model to address.

Two examples of these discrepancies are shown in Figure 8. Along the New Jersey bank of the Hudson River (Figure 8A), two overpasses for I-78 are accounted for in the model's lidar-derived DEM, but do not allow for flow of water through the underpass.

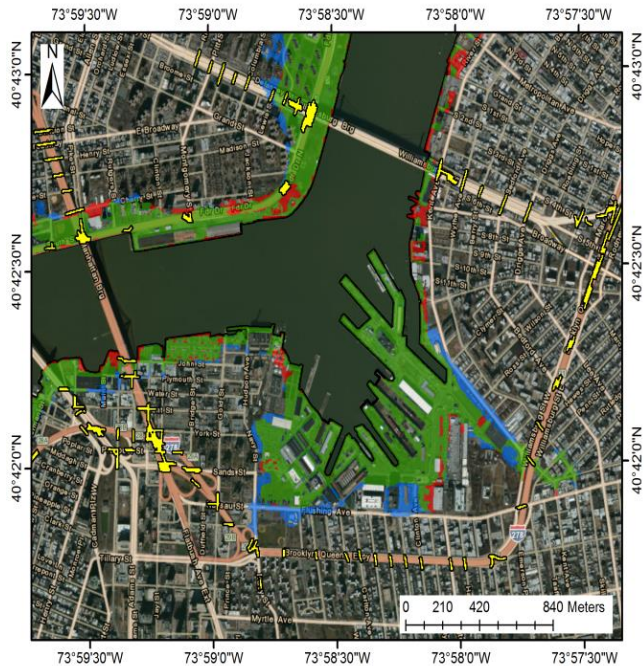

**Figure 9.** Area comparison results after underpass inclusion, with highway underpasses shown in yellow for Brooklyn near the Navy Yard. Of the 66 underpasses shown, 15 (22.7%) are within the storm surge impact area of Hurricane Sandy.

Thus, the model under-predicts flooding along Thomas McGovern Drive by as much as 258.6m (Figure 6D), and this discrepancy adversely affected the distance and area comparisons (Table 3 and Table 4). Similar roadway infrastructure issues with the DEM cause inundation along Kent Avenue to be blocked by an overpass for I-278. This caused the model to under-predict flooding east of the overpass by 169.4m (Figure 6A), and over-predict flooding west of the overpass (Figure 8C). If we account for highway underpasses within the model sub-grid, the impact of physical impediments for fluid flow not accounted for in the model's DEM may be minimized.

*5.4 Augmented results with resolved underpasses*

As demonstrated in the high-resolution simulation results, accurate representation of urban infrastructure in a hydrodynamic model's DEM is vital to predictive accuracy, because it uniquely affects the fluid flux through each grid cell side, which ultimately determines the water depth and extent of flooding via distribution of water volume within each grid cell. The model results shown in Figure 8 highlight two unique locations where the model failed to accurately spatially predict urban inundation due to misrepresentation of dense urban infrastructure in the lidar-derived DEM. Table 5 reviews these two locations using model output edge porosity metrics and partial wetting and drying cell area statistics before and after consideration of urban and highway infrastructure in the DEM used to construct the model sub-grid.

In an effort to improve upon the spatial comparison results reported in the previous sections, streets for New York City and New Jersey were retrieved from Open Street Map, merged together, and clipped to the study area. Bridges were extracted to a separate layer, and the merged streets were split using 50m intervals. Selected by location, 50m segments from the merged streets layer crossed by the extracted bridges layer were exported to a new layer comprised of 50m segments containing street underpasses. The street underpass layer was subsequently buffered by 8m to account for average conventional street widths and walkways, dissolving all resulting attribute fields and overlapping layers into one. Subsequently, the explode multipart function was utilized to separate unique individual underpass segments, and the FID value was copied into the ID field before executing the zonal minimum function referencing the original DEM without highway underpass consideration to overwrite the original elevations in the DEM with the average elevations for the 50m underpass segments. Cell statistics were calculated using the following settings: extent = union, snapping = original top, and cell size = minimum, to conveniently account for all occluded underpasses within the study region as shown in the example in Figure 9. The sub-grid simulation was rerun with the updated DEM including highway underpasses, and significantly improved the capability for the sub-grid model to match overall distance error metrics with a mean absolute distance difference of 32m, a 15.8% improvement (Table 3). Area comparisons improved to an 80% spatial match (Table 4), all due to minimal additional effort to account for urban highway underpasses during DEM preparation.

Of the 810 areas with highway underpasses detected within the study area, only 238 (29.4%) were affected by Hurricane Sandy's storm surge. Hurricane Sandy was a category 1 storm at landfall. Stronger storms with greater maximum flood extents such as the category 3 New England Hurricane of 1938 with 125mph winds and a reported maximum storm surge height of 5m (18ft.) in Long Island, would observe a higher relevance ratio for overpass importance on spatial results. The benefit of this quick overpass adjustment approach is that it not only has the capacity to improve storm surge model results in urban environments, but the method is not storm specific, with broader relevance to future storms and essentially universally applicable to all urban coastal regions with elevated roadway infrastructure.

The bridge overpass at Thomas McGovern Dr. and I-278 in Bayonne, NJ, featured in Figure 8A initially blocked storm surge inundation yielding zero porosity at grid edge #2134 (Table 5). After resolving the highway underpasses in the model DEM and rebuilding the sub-grid, a new time-varying porosity of 0.50 was measured during the peak inundation period with a water elevation of 3.48m above NAVD88. This indicates that 50% of grid edge #2134 is wet during the storm surge peak, now effectively flooding the lower elevation neighborhood in Bergen/Lafayette on the other side of I-278, in better agreement with the FEMA maximum inundation extents (Figure 8B). Additionally, grid cell #10471 was initially completely dry during peak storm surge, but after rectification of the bridge underpass elevations for free fluid flow, a maximum wet area of 26,800m² (67%) of the 200×200m computational grid cell was observed. Similarly, the bridge overpass at Kent Ave. and I-78 featured in Figure 8C originally resulted in zero porosity at grid edge #12901 before highway underpasses were included in the model DEM. Once accounted for, a new peak porosity of 0.25 was observed with a water elevation of 3.41m above NAVD88, revealing that 25% of grid edge #12901 is wet, yielding a maximum wet area of 20,550m² (51%) in grid cell #6644 during the storm surge peak (Table 5; Figure 8D). Partial wetting and drying is effectively illustrated in the three-dimensional depiction of the maximum flooding modeled along Kent Ave. in Figure 10, in conjunction with the time-varying maximum wet area metrics presented in Table 5.

## 6. Discussion and Conclusion

An important aspect of this study is the emphasis on the use of high resolution coastal DEMs and bathymetric sounding measurements combined with sub-grid modeling to provide accurate storm surge and inundation results. In evaluating the model accuracy, high water marks and spatial inundation maps produced by the USGS and FEMA, respectively, were used extensively to verify modeled storm surge heights and inundation extents. The USGS-collected high-water mark data observations were reported to be vertically accurate within 0.079m at the 95% confidence level, and within 3m horizontally, with errors originating from datum differences and surveyed uncertainty (McCallum at al., 2013). Despite the considerable efforts made by the USGS and FEMA, the datasets are not perfect.

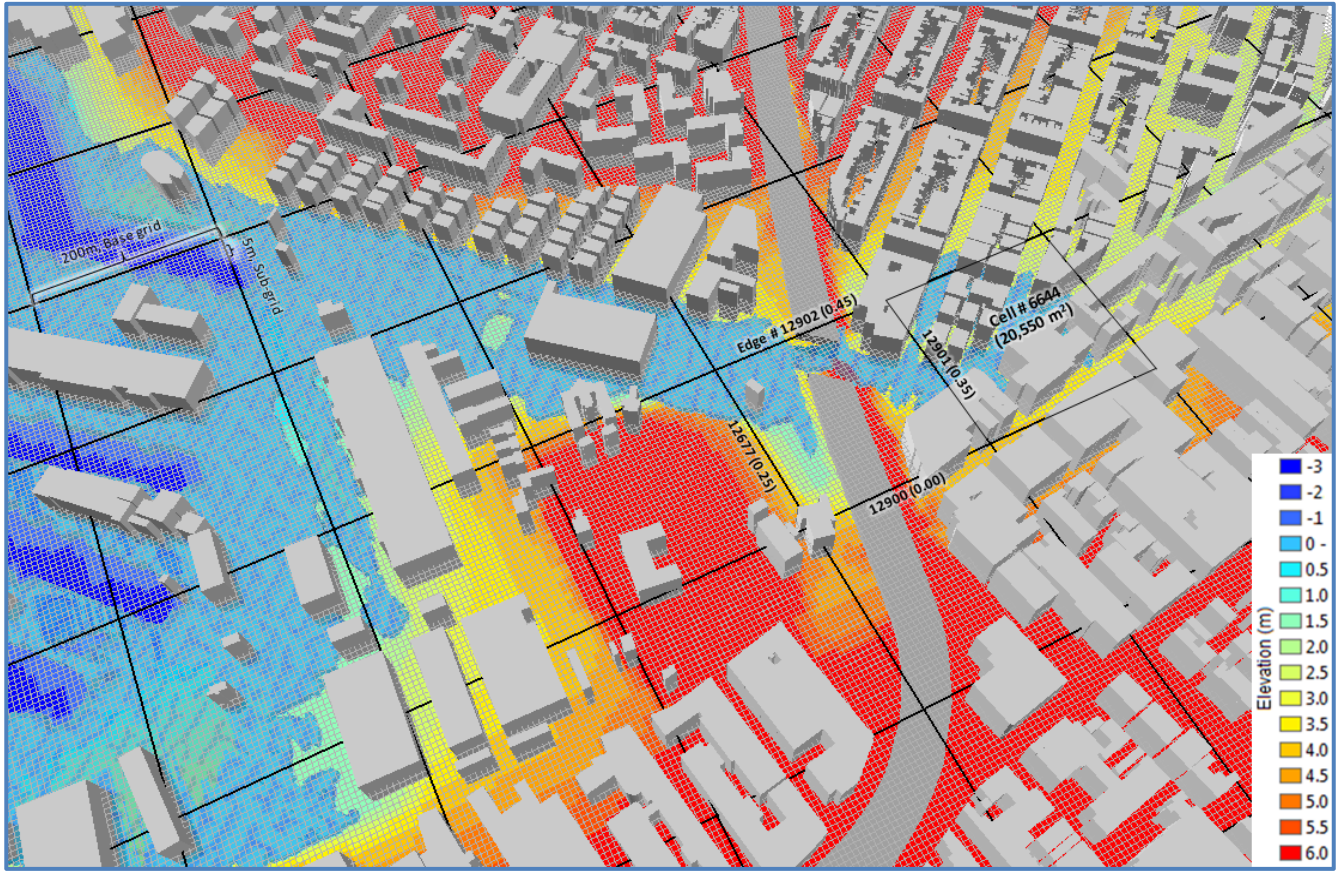

**Figure 10.** Three-dimensional depiction of the maximum flooding extents along Kent Ave. with buildings and overpasses extruded to show detail in Brooklyn during Hurricane Sandy. Sub-grid topography and bathymetry elevations (5m resolution) are shown with corresponding color ramp within each black outlined 200m base grid cell. The Kent Ave. underpass elevations have been corrected in the lidar-derived DEM to accurately simulate flooding on the other side of I-278.

The high water marks are inherently inhomogeneous and sparse. The high water marks were presumably linearly interpolated to provide a maximum inundation extent map, which can lead to erroneous estimation of the water surface elevation in areas far from or lacking nearby surveyed source data (Schmid et al., 2014). Also, this flood map was produced via intersection of the interpolated water elevations recorded at the high water marks with the best available DEM to produce a data-derived bathtub model of GIS inundation extents (FEMA MOTF, 2013). The bathtub model is data-driven, and does not account for the conservation of mass, effect of water velocities, and wind stress, all potentially important parameters during hurricane events.

**Table 5**. Grid edge porosity metrics before and after consideration of infrastructure in the DEM used to construct the model sub-grid.

| Data Location | Edge Porosity | | | |
|---|---|---|---|---|
| | Grid Edge # | Peak Water Elevation (m) | Max. Porosity without Underpasses and Buildings | Max. Porosity with Underpasses and Buildings |
| T. McGovern Dr. (west) | 2134 | 3.48 | 0.00 | 0.50 |
| T. McGovern Dr. (north) | 2294 | 3.48 | 0.00 | 0.25 |
| T. McGovern Dr. (south) | 2292 | 3.48 | 0.50 | 0.50 |
| T. McGovern Dr. (east) | 2293 | 3.48 | 0.80 | 0.80 |
| Kent Ave. (west) | 12677 | 3.41 | 0.35 | 0.25 |
| Kent Ave. (north) | 12902 | 3.41 | 0.40 | 0.45 |
| Kent Ave. (south) | 12900 | 3.41 | 0.00 | 0.00 |
| Kent Ave. (east) | 12901 | 3.41 | 0.00 | 0.25 |
| | Partial Wetting and Drying | | | |
| | Grid Cell # | Peak Water Elevation (m) | Max. Wet Area without Underpasses and Buildings (m²) | Max. Wet Area with Underpasses and Buildings (m²) |
| T. McGovern Dr. / I-278 | 10471 | 3.48 | 0 | 26,800 |
| Kent Ave. / I-78 | 6644 | 3.41 | 0 | 20,550 |

The numerical sub-grid model used in this study contains the conventionally accepted assumptions of hydrodynamic and sub-grid models addressed in Wang et al., 2014. The sub-grid model setup described in this paper does not consider the rainfall experienced in New York City during Hurricane Sandy. The cumulative precipitation measured from October 29 - 31 in Central Park was 0.95in, while 0.56in was recorded at JFK International Airport (NOAA NCDC, 2012). Since a majority of reported flooding during the storm was attributed to the hurricane's considerable storm surge, smaller sources attributed to precipitation may be negligible. Likewise, the effects of infiltration through ground surfaces and subsurface drainage sinks, including storm drains diverting water to reservoirs, retention ponds, or back to the estuaries by way of storm drains and runoff were not considered in the model. The reason that the precipitation and subsurface transport were not considered is because these two processes are rather complicated in an urban area and the technologies for properly representing them are still lacking. Nevertheless, in terms of prediction of the inundation peak, we feel confident that the major source of water from the storm tide during Hurricane Sandy was dominant, with precipitation and drainage playing secondary roles in the inundation.

In all, this research was performed to demonstrate a new method to more effectively predict spatial storm surge extents in urban environments than earlier models which shy away from directly addressing the built environment. Upon determining originations of model uncertainty to be minimal, this leads to the conclusion that the newly developed multi-scale sub-grid model was effectively tuned for urban environments. This was accomplished by utilizing lidar-derived topography, high-resolution bathymetry, and building structures in New York City during Hurricane Sandy. Sub-grid modeling enabled large-area utilization of high-resolution topography and bathymetry data in ways that would be impractical using conventional hydrodynamic modeling approaches bereft of partial wetting and drying capabilities. Thus, modelers can use large, dense lidar datasets with advanced sub-grid modeling techniques, which operates across scales, computing hydrodynamics on a coarse computational scale with local topography being stored at sub-grid scales. Street-level sub-grid model performance was assessed via spatial inundation comparison with verified field measurements using USGS-collected high water marks, FEMA-collected data regarding inundated schools, and calculated area and distance differentials using FEMA's maximum extent of inundation map.

Spatial verification of the inundation depths predicted by the sub-grid model were addressed via comparison with 73 high water mark measurements collected by the USGS and by 80 FEMA-reported water level thicknesses at inundated schools throughout the sub-grid domain separated by state. Average statistics for the 73 USGS-recorded high water marks for New York and New Jersey were: 0.120m and 0.347m for root-mean-squared error, respectively. Aggregated statistical metrics for the 80 FEMA-reported inundated schools for root-mean-squared error in New York were 0.3293m and New Jersey were 0.4760m. Larger differences and errors reported in the point-to-point comparisons for New Jersey relative to New York were largely due to the lack of building representation in the sub-grid DEM for the New Jersey side of the Hudson River, and were a significant indication that the representation of buildings as a physical impediment to fluid flow is critical to urban inundation modeling.

Maximum spatial extent of inundation was assessed using FEMA's spatial flood coverage map to calculate distances between the model's predicted maximum flood extent and FEMA's reported flood maximums, wherein the sub-grid model had an absolute mean distance difference of nearly 40m (38.43m) or nearly eight 5m-resolution sub-grid pixels. The initial spatial verification calculated a difference map to successfully conduct a complete area comparison, which resulted in a 75.15% spatial match with 11.41% area model over-prediction and 13.44% model under-prediction. Incorporation of highway underpasses significantly improved the capability for the sub-grid model to match overall distance error metrics for absolute mean difference from 38m to 32m (a 15.8% improvement) and area comparisons from 75% spatial match to 80% with minimal additional effort during the DEM preparation phase.

**Acknowledgements**

The authors would like to acknowledge the NOAA Tides and Currents database for useful meteorological observations, and water level predictions and observations used for boundary conditions, calibrations, and forcing the model.